\documentclass[aps, amsfonts, amsmath, amssymb, showpacs,showkeys,onecolumn,preprint,preprintnumbers]{revtex4}
\usepackage{amssymb}
\usepackage{amsmath}%
\usepackage{amsfonts}%
\usepackage{slashbox}
\usepackage{mdframed}
\usepackage{hyperref}

\newtheorem{theorem}{Theorem}[section]

\newtheorem{lemma}[theorem]{Lemma}

\begin{document}

\title[Remarkable New Identity]{A Remarkable New Identity Satisfied by the Dirac Matrices of a Bilocal Field Theory}

\author{Patrick L. Nash}
\email
{Patrick299Nash@gmail.com}

%

\affiliation
{
Department of Physics and Astronomy, Retired\\
The University of Texas at San Antonio\\
San Antonio, Texas 78249-0697
}

\date{\today}

\begin{abstract}
In 1925  Elie Cartan described `triality' \cite{CARTAN25}, \cite{CARTAN}
as a symmetry between SO$(8; \mathbb{C})$ vectors and the  two types of  Spin$(8; \mathbb{C})$ spinor.
It is known that the reduced generators of the Clifford algebra $\mathbb{C}_{8}$
defined on
the real, eight-dimensional Euclidean space $\mathbb{E}_{8}$ satisfy an identity that
guarantees the existence of
matrix  representations (acting on the vector and spinor bundles of $\mathbb{E}_{8}$) of triality.

Analogously, let $\mathbb{E}_{4,4}$
denote a real eight-dimensional pseudo-Euclidean vector space that is endowed with
an indefinite inner product with signature
$(+,+,+,-\,;\,-,-,-,+)$.
As a normed vector space, $\mathbb{E}_{4,4} \cong M_{3,1} \times {}^{*}\!M_{3,1}$,
where $M_{3,1}$ and ${}^{*}\!M_{3,1}$ denote real four-dimensional Minkowski spacetimes,
with opposite signatures.
The reduced generators (i.e., the Dirac matrices) of the pseudo Clifford algebra $\mathbb{C}_{4,4}$
defined on $\mathbb{E}_{4,4}$ satisfy an identity $\,$ \cite{NASH86} $\,,\,$  \cite{NASH90}
that guarantees the existence of invertible linear mappings  between each of the two
types of $\overline{S0(4,4; \mathbb{R})}$ spinor and the ${S0(4,4; \mathbb{R})}$ vector,
thereby realizing matrix  representations  of triality
that act on the  vector and spinor bundles of the spacetime $\mathbb{E}_{4,4}$.

In this note we generalize this identity (see Eq.[\ref{newIdentity}]).
\end{abstract}


\maketitle

\section{Introduction and Notation}
In 1925  Elie Cartan described `triality' \cite{CARTAN25}, \cite{CARTAN}
as a symmetry between three types of geometrical objects that may be defined on
real, eight-dimensional $\mathbb{R}^{8}$
and transform linearly under either SO$(8; \mathbb{C})$ or Spin$(8; \mathbb{C})$,
namely  a symmetry between SO$(8; \mathbb{C})$ vectors and the  two types of  Spin$(8; \mathbb{C})$ spinor (semi-spinors
of the first type and semi-spinors of the second type, in the
terminology of Cartan).

Analogously, let $\mathbb{E}_{4,4}$
denote a real eight-dimensional pseudo-Euclidean vector space that is endowed with
an indefinite inner product with signature
$(+,+,+,-\,;\,-,-,-,+)$ (see Gray \cite{Gray:1969}).
As a normed vector space, $\mathbb{E}_{4,4} \cong M_{3,1} \times {}^{*}\!M_{3,1}$, where
$M_{3,1}$ denotes a real four-dimensional Minkowski spacetime manifold that is endowed with the pseudo-Euclidean metric
$\eta_{3,1}$ = diag(1, 1, 1, -1), and
$ {}^{*}\!M_{3,1}$ denotes a real four-dimensional Minkowski spacetime that is endowed with the pseudo-Euclidean metric\\
diag(-1, -1, -1, 1) = $-\eta_{3,1}$.
$M_{3,1} \times {}^{*}\!M_{3,1}$ may be regarded as
a classical phase space of a single relativistic point particle,
or a spacetime that carries a bilocal Minkowski field theory
(appropriate restrictions on the automorphism groups of
$\mathbb{E}_{4,4} \cong M_{3,1} \times {}^{*}\!M_{3,1}$ are implied).

The reduced generators (i.e., the Dirac matrices) of the pseudo Clifford algebra $\mathbb{C}_{4,4}$
defined on $\mathbb{E}_{4,4}$ satisfy an identity $\,$ \cite{NASH86} $\,,\,$   \cite{NASH90}
that guarantees the existence of invertible linear mappings  between each of the two
types of $\overline{S0(4,4; \mathbb{R})}$ spinor and the ${S0(4,4; \mathbb{R})}$ vector,
thereby realizing matrix  representations  of triality
that act on the  vector and spinor bundles of the spacetime $\mathbb{E}_{4,4}$.
%
%
In this note we generalize this remarkable identity Eq.[\ref{uid}]
to Eq.[\ref{newIdentity}].
Simple applications of this formalism
are given in Sections [\ref{Application2}] and [\ref{Application1}].



$\mathbb{E}_{4,4}$ is an orientable  differentiable manifold that, of course, admits a global, right-handed Cartesian atlas
(as well as many other ``curvilinear" and general coordinate systems).
Let  $\mathbf{x} \in \mathbb{E}_{4,4}$ and let the 8 scalars
$x^A \in \mathbb{R}$, $\;A, B, ... = 1, 2, ... , 8$,
denote the Cartesian coordinates of $\mathbf{x}$ with respect to a  global, right-handed Cartesian atlas.
Let $T_x(\mathbb{E}_{4,4})$  denote the tangent space at $\mathbf{x}$.
$T_x(\mathbb{E}_{4,4})$ is isomorphic to $\mathbb{E}_{4,4}$.
The right-handed frame
$
\left\{
\frac{\partial}{\partial\,x^A}\,:\,A = 1,\ldots ,8
\right\}
$
that is adapted to these coordinates
is orthogonal and pseudo-normal with respect to the metric defined below,
and comprises a basis of  $T_x(\mathbb{E}_{4,4})$.
This coordinate system 
and frame are simply called a ``canonical frame".
A vector field $V$ at x, $V_x = V^A(x)\,\frac{\partial}{\partial\,x^A}\,\in\,T_x(\mathbb{E}_{4,4})$, has contravariant components
$V^A(x)$ with respect to a canonical frame.
Here the A, B, ... = 1, ... , 8 are to regarded as $T_x(\mathbb{E}_{4,4})$ vector indices,
and not as indices that enumerate the scalars $x^A$; the interpretation of an index should always be clear from context.

\section{Dirac Matrices on $\mathbb{E}_{4,4}$}
\subsection{Representations of  SO$(8; \mathbb{C})$}

There is  a well known
relationship between Clifford algebras $C_{n}$
and the spinor representations of the classical complex orthogonal groups;
see, for example,
Boerner, \emph{The Representations of Groups} \cite{BOERNER}.
In particular, the Clifford algebra $C_8$ may be defined as the algebra
generated by a set of eight elements $e_j, j, k\, = 1,\ldots,8,$ that anticommute with each
other and have unit square
$e_j \, e_k + e_k \, e_j = 2 \, \delta_{j\,k} \, \mathbb{I}_{16 \times 16}$,
where $\mathbb{I}_{16 \times 16} = $
${16 \times 16}$ unit matrix.
The scaled commutators
$\frac{1}{4}\left(e_j \, e_k - e_k \, e_j\right)$
computed from an irreducible 16-dimensional representation of the $e_j$
are the infinitesimal generators of
a reducible 16-dimensional representation of Spin$(8; \mathbb{C})$,
which is the universal double covering
of the special orthogonal group SO$(8; \mathbb{C})$.
This  16-dimensional representation of is fully reducible to the direct sum of two
inequivalent irreducible $8 \times 8$ spin representations
of the infinitesimal generators of
Spin$(8; \mathbb{C})$
\cite{CARTAN25},
\cite{CARTAN},
\cite{Springer:582165},
\cite{Harevy1990:9780123296504},
\cite{Baez}.
The fundamental irreducible  vector representation of SO$(8; \mathbb{C})$ is also $8 \times 8$.
The Dynkin diagram for $D_4 \cong \textrm{SO}(8; \mathbb{C}) $
is symmetrical and pictured in Figure 1.
The central node corresponds to the adjoint representation.
The three outer nodes correspond to the vector representation (left-most node),
type 1 spinor and type 2 spinor
representations of Spin$(8; \mathbb{C})$.
The ``left-handed"  and ``right-handed" Spin$(8; \mathbb{C})$ spinors have $\overline{S0(4,4; \mathbb{R})}$ counter parts that
are denoted  $\psi_{(1)}$ and $\psi_{(2)}$ in this paper,
and transform, respectively, under
two inequivalent
real $8 \times 8$ irreducible  spinor representations  of
$\overline{S0(4,4; \mathbb{R})}$ that we have called  ${D_{(1)}}$ (type 1) and ${D_{(2)}}$ (type 2).

\begin{figure}
\label{D4}
\begin{picture}(300,20)(0,0)
\put(32,0){\circle{6}}
\put(35,0){\line(1,0){25}}
\put(63,0){\circle{6}}
\put(65,3){\line(1,1){17}}
\put(84, 22){\circle{6}}
\put(65,-3){\line(1,-1){17}}
\put(84,-22){\circle{6}}
\end{picture}
\vspace{0.4cm}
\caption{Dynkin diagram for $D_4$}
\end{figure}
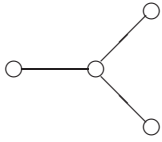
\vspace{0.5cm}

$SO(4,4; \mathbb{R})$ is a real form of the classical complex orthogonal group  SO$(8, \mathbb{C})$.
$O(4,4; \mathbb{R})$ (respectively,  $SO(4,4; \mathbb{R})$)
may be defined as the group of all real matrices   (respectively, with unit determinant)
that preserve the
norm squared of  $V_x\,\in\,T_x(\mathbb{E}_{4,4})$,
which is the quadratic form
\begin{displaymath}
(V^8)^2 +(V^1)^2 + (V^2)^2 + (V^3)^2 - \left[ (V^4)^2 + (V^5)^2 + (V^6)^2 + (V^7)^2 \right].
\end{displaymath}
$O(4,4; \mathbb{R})$ is a pseudo-orthogonal Lie group that possess two
connected components \cite{BOERNER},\cite{HELGASON},
with $SO(4,4; \mathbb{R})$ being the identity component (the connected component containing the identity matrix).
Spin$(4,4,\mathbb{R})$,
alternatively denoted
$
\overline{SO(4,4; \mathbb{R})}
$,
is the 2-to-1 covering group of $SO(4,4; \mathbb{R})$.

$\mathbb{E}_{4,4}$ may be endowed with both
${SO(4,4; \mathbb{R})}$-invariant and
$\overline{SO(4,4; \mathbb{R})}$-invariant pseudo-Euclidean
metrics that may each be represented in terms of an $8 \times 8$ matrix with real matrix elements.

In a canonical $\mathbb{E}_{4,4}$ frame the ${SO(4,4; \mathbb{R})}$-invariant pseudo-Euclidean
metric tensor $\mathbb{G}$ (respectively,  inverse  ${\mathbb{G}}^{-1}$)
has components $\mathbb{G}_{A\,B}$
(respectively,  $\left({{\mathbb{G}}^{-1}}\right)^{A\,B} = \mathbb{G}^{B\,A} = \mathbb{G}^{A\,B}$)
that are given by
\begin{eqnarray}
\mathbb{G}_{A\,B} = \mathbb{G}^{A\,B}
&=&
\left(
\begin{array}{cc}
 \eta_{3,1} & 0\\
 0 &  -\eta_{3,1}
\end{array}
\right)
\end{eqnarray}
The indefinite inner product is realized as
$T_x(\mathbb{E}_{4,4}) \times T_x(\mathbb{E}_{4,4}) \ni (V_x,{V'}_x) \;$
$\mapsto $
$\; < V_x, V'_x> = \mathbb{G}_{A\,B} V_x^A {V'}_x^B \in \mathbb{R}$.
%

\subsection{Spinor representations of  $\overline{S0(4,4; \mathbb{R})}$}

There exist  two inequivalent real $\overline{S0(4,4; \mathbb{R})}$
basic 8-component spinor representations of
${S0(4,4; \mathbb{R})}$.
They are defined in Eqs.[\ref{reps}] and simply denoted as ${D_{(1)}}$ (type 1)  and ${D_{(2)}}$ (type 2).
The  $\overline{S0(4,4; \mathbb{R})}$ invariant metric, denoted
$\sigma$, is invariant under the action of both ${D_{(1)}}$ and ${D_{(2)}}$.
Let $S_x^{(j)}(\mathbb{E}_{4,4})$,  $\;j = 1, 2$,  denote the two
distinct basic real 8-component spinor vector spaces at $\mathbf{x}$,
endowed with respective automorphism groups  ${D_{(j)}}$.
As vector spaces each  is isomorphic to $\mathbb{E}_{4,4}$.
%
(Thus, as vector spaces, both of the $S_x^{(j)}(\mathbb{E}_{4,4})$ and $T_x(\mathbb{E}_{4,4})$
are each isomorphic to $\mathbb{E}_{4,4}$
but with different automorphism groups.)
A spinor element  $\psi_{(j)} \in S_x^{(j)}(\mathbb{E}_{4,4})$ has components $\psi_{(j)} ^a \in \mathbb{R}$.
In this note, for simplicity, we do not distinguish the spinor index  on $\psi_{(1)}$ from
that on $\psi_{(2)}$
[using a convention such as
${\psi_{(1)}}{}_{a}$ and ${\psi_{(2)}}^{\dot{a}}$ for spinor components, for example].
%
%
%

The disjoint union of tangent spaces $T_x(\mathbb{E}_{4,4})$ at all points  $\mathbf{x} \in \mathbb{E}_{4,4}$
gives the  $SO(4,4; \mathbb{R})$ tangent bundle $T(\mathbb{E}_{4,4})$  over  $\mathbb{E}_{4,4}$.
In this case, it is a trivial bundle
$\mathbb{E}_{4,4} \times T_x(\mathbb{E}_{4,4}) \,  \stackrel{\pi}{ \rightarrow} \, \mathbb{E}_{4,4}$,
with the natural  projection $\pi$ of the first factor in the Cartesian product.
Clearly there also exist two distinct trivial 16-dimensional real basic 8-component spinor  bundles
$S^{(1)}(\mathbb{E}_{4,4})$ and  $S^{(2)}(\mathbb{E}_{4,4})$,
each with base space $\mathbb{E}_{4,4}$
but with fibers   $S_x^{(1)}(\mathbb{E}_{4,4})$ and  $S_x^{(2)}(\mathbb{E}_{4,4})$, respectively.
For each of the three bundles we denote the natural  projection of the first factor in the Cartesian product by  $\pi$.

$\mathbb{E}_{4,4}$  may  be endowed with a $\overline{S0(4,4; \mathbb{R})}$ invariant metric
$\sigma$ $\,$ \cite{NASH86} $\,,\,$ \cite{NASH90}
that we represent as
\begin{eqnarray}
\sigma
&=&
\left(
\begin{array}{cc}
 0 & 1\\
 1 &  0
\end{array}
\right)
\label{sigma}
\end{eqnarray}
where 0 denotes the 4 x 4 zero matrix and 1 denotes the 4 x 4 unit matrix.
The
matrix elements of $\sigma$ are denoted $\sigma_{a b}$ = $\sigma_{b a}$, where $a, b, \ldots = 1, \ldots ,\,8$ are
$\overline{S0(4,4; \mathbb{R})}$ spinor indices (elaborated in Eqs.[\ref{reps}] through [\ref{L}] below).
Note that $\sigma^2$ is equal to the unit matrix, so that the eigenvalues of $\sigma$ of are $\pm 1$.
Since the trace of  $\sigma$ is zero, these eigenvalues occur with equal multiplicity.

The $\overline{S0(4,4; \mathbb{R})}$ invariant (pseudo) norm-squared $\|\psi_{(j)}\|^2$
of basic real 8-component spinors $\psi_{(j)} \in S_x^{(j)}(\mathbb{E}_{4,4})$
is the $\overline{S0(4,4; \mathbb{R})}$-invariant quadratic form
$\;\;\psi_{(j)}^a \;\, \sigma_{a b} \;\,\psi_{(j)}^b$.
We define an oriented spinor basis $e_a$ of  $S_x^{(1)}(\mathbb{E}_{4,4})$
normalized according to
\begin{equation}
\label{neaeb}
< e_a, e_b > = \sigma_{a b}
\end{equation}
(the oriented spinor basis of  $S^{(2)}(\mathbb{E}_{4,4})$ also satisfies Eq.[\ref{neaeb}]),
so that\\
$< \psi_{(1)}, \psi_{(1)} > = < \psi_{(1)}^a \;e_a, \psi_{(1)}^b \;e_b >  $
=$\;\;\psi_{(1)}^a \;\, \sigma_{a b} \;\,\psi_{(1)}^b$
=$\widetilde{\psi}_{(1)}\;\sigma\;\psi_{(1)}$,
where the tilde denotes transpose.
For brevity we employ the shorthand
$\mathbf{u} \in S^{(1)}(\mathbb{E}_{4,4})$ and
$u^a \in S^{(1)}(\mathbb{E}_{4,4})$ to denote
$e_a\,u^a \in S^{(1)}(\mathbb{E}_{4,4})$,
with similar conventions implied for $T(\mathbb{E}_{4,4})$ and $S^{(2)}(\mathbb{E}_{4,4})$.

We also define  an oriented vector basis $\epsilon_A$
of $T_x(\mathbb{E}_{4,4})$
normalized according to
\begin{equation}
\label{neAeB}
< \epsilon_A, \epsilon_B > = \mathbb{G}_{A\,B}.
\end{equation}
These two sets of basis vectors are related by Eq.[\ref{eAea}] below.


The basic spinor representation of the pseudo-orthogonal
group $SO(4,4; \mathbb{R})$ may be constructed from the irreducible generators
$t^A$,  A = 1, ... , 8 , of the pseudo-Clifford algebra $C_{4,4}$ $\,\,$ \cite{BOERNER}$\,,\,$\cite{BRAUER}$\,,\,$ \cite{Lord68}.
Following Brauer and Weyl  we  call
such irreducible $C_{2n-2}$ generators ``reduced Brauer-Weyl generators" $\;$ \cite{BRAUER}.
We begin the construction of
a representation of $\overline{S0(4,4; \mathbb{R})}$ by defining eight
real $8 \times 8$ matrix reduced
Brauer-Weyl generators $ \tau^A, \overline{\tau}^A $,
A,B, ... = 1,...,8 , of the pseudo-Clifford algebra $C_{4,4}$ that anticommute
and have square $\pm 1$.
(The  $\overline{\tau}^A $ matrices play the role of the Dirac Matrices on $\mathbb{E}_{4,4}$.)
We realize this by requiring that the tau matrices satisfy (the tilde denotes transpose)

\begin{equation}
\sigma \overline{\tau}^A = {\widetilde{ \sigma\tau^A } }
= {\widetilde{ \tau } }^A \sigma
\label{tau-bar}
\end{equation}
and
\begin{equation}
\tau^A \overline{\tau}^B + \tau^B \overline{\tau}^A = 2\,  \mathbb{I}_{8 \times 8} \; \mathbb{G}^{A\,B} =
\overline{\tau}^A \tau^B + \overline{\tau}^B \tau^A
\label{tau-tau},
\end{equation}
where $ \mathbb{I}_{8 \times 8} $ denotes the $8 \times 8$ unit matrix.
Denoting the matrix elements of
$\tau^A$ by
$\tau^{A\;a}_{\phantom{A\; ab} b}$, we may write Eq.[\ref{tau-bar}] as
\begin{equation}
\overline{\tau}^{A}_{\phantom{A}a b} = \tau^{A}_{\phantom{A}b a},
\label{tau-ab}
\end{equation}
where we have used  $\sigma$  to lower the spinor indices. In general,  $\sigma$  (respectively,  $\sigma^{-1}$ )
will be employed to lower (respectively, raise) lower case Latin indices (i.e.
a $\overline{S0(4,4; \mathbb{R})}$ spinor index of either type).

The following identity is occasionally useful.
Let
$\psi  \in \, S^{(1)}(\mathbb{E}_{4,4})$ be an arbitrary real eight component type-1 spinor field
(a section of the type-1 spinor bundle $S^{(1)}(\mathbb{E}_{4,4})$).
Consider

\begin{eqnarray}
\label{tautauID}
\widetilde{\psi} \, \sigma \,
\overline{\tau}^{A}\,
\tau_{B} \, \psi
\,&=&\,
\left(
\widetilde{\psi} \, \sigma \,
\overline{\tau}^{A}\,
\tau_{B} \, \psi
\right)^{T}
\,
=\,
\left(
\widetilde{\psi} \, \sigma \,
\overline{\tau}_{B}\,
\tau^{A} \, \psi
\right)
\;\;
\textrm{ , by  Eq.[\ref{tau-bar}]}
\nonumber \\
&=&
\frac{1}{2}
\widetilde{\psi} \, \sigma \,
\left(
\overline{\tau}^{A}\,
\tau_{B}\;+\;
\overline{\tau}_{B}\,
\tau^{A} \,
\right)
\, \psi
\nonumber \\
&=&
\delta_{B}^{A}\,
\widetilde{\psi} \, \sigma \, \psi
\;\;
\textrm{ , using  Eq.[\ref{tau-tau}]}.
\end{eqnarray}


We adopt a real irreducible $8 \times 8$ matrix representation of the tau
matrices that is adapted to the $X^8$-axis, in which
$\overline{\tau}^{8} =  \mathbb{I}_{8 \times 8}  = \tau^{8}$.
Then, by Eq.[\ref{tau-tau}],
$\overline{\tau}^{A} = - \tau^{A}$ for A = 1, ... ,7. Hence, again
by  Eq.[\ref{tau-tau}],
$(\tau^{A})^2$ is equal to $- \mathbb{I}_{8 \times 8} $  for A = 1,2,3 and is equal to  $+ \mathbb{I}_{8 \times 8} $ for A = 4,5,6, 7, 8.
The Appendix displays one possible representation.

\section{The New Identity}
The tau matrices verify an important identity $\,$ \cite{NASH86} $\,,\,$ \cite{NASH90} that encodes triality:
Let M be any $8 \times 8$ matrix satisfying (recall that the tilde denotes transpose)
\begin{equation}
\widetilde{\sigma \,M} = \sigma \, M
\label{sU}
\end{equation}
(i.e., $\sigma M$ is a symmetric matrix)
and moreover transforming under $\overline{S0(4,4; \mathbb{R})}$ according to
\begin{equation}
M \mapsto {D_{(1)}} M {D_{(1)}}^{-1}
\label{U}
\end{equation}
(see Eq.[\ref{D1}], below).
Then $\,$ \cite{NASH86}${}^{,}$ \cite{NASH90}
\begin{equation}
\tau_A M \, \overline{\tau}^A =  \mathbb{I}_{8 \times 8}  \; tr(M),
\label{uid}
\end{equation}
where, as above, $ \mathbb{I}_{8 \times 8} $ denotes the $8 \times 8$ unit matrix.
This is a  remarkable identity because this linear combination of eight terms involving an arbitrary
real $8 \times 8$ symmetric matrix $\sigma   M$ is proportional to the unit matrix,
and there are 36 linearly independent real $8 \times 8$ matrices $M$ such that $\sigma  M$ is a symmetric matrix
(these are given below in Eq.[\ref{symmetricM}]).

This is a special case of another  simple, but also remarkable, general identity
that we record as
\begin{theorem}
\label{tj}
Let $M$ be an arbitrary $8 \times 8$ matrix that transforms under $\overline{S0(4,4; \mathbb{R})}$ according to
$M \mapsto {D_{(1)}} M {D_{(1)}}^{-1}\;
$.
$M$  has matrix elements  $M^{a}_{\phantom{a }b}$.
Note that
$M \, - \, \sigma ^{-1} \, \widetilde{\left(\sigma \, M\right)}$
is twice
$\sigma^{-1} $ times the anti-symmetric part of
$\sigma  M$.
The generalization of Eq.[\ref{uid}] is
\begin{equation}
\label{newIdentity0}
\tau _{ (\mu) }\,M\, \overline{\tau}^{ (\mu) }
\,=\,
\,-\, \tau _{ (\mu) } \, \textrm{tr}\left(\overline{\tau}^{ (\mu) } \, M\right)
\,+\,
2 \left(\,\mathbb{I}_{8\times8} \, \textrm{tr}\left( M \right)+ \,
M \, - \, \sigma ^{-1} \, \widetilde{\left(\sigma \, M\right)}\,\right)
\end{equation}
or
\begin{equation}
\label{newIdentity}
\left(\tau _{ (\mu) }\right)_{\phantom{a }b}^a \left( \overline{\tau}^{ (\mu) } \right)_{\phantom{c }d}^c
\,=\,
\,-\,\left(\tau _{ (\mu) }\right)_{\phantom{c }d}^a \left(\overline{\tau}^{ (\mu) }\right)_{\phantom{a }b}^c
\,+\,
2 \left(\delta _b^a \delta _d^c+\delta _d^a \delta _b^c-\sigma ^{a c} \sigma _{b d}\right)
\end{equation}

\end{theorem}

The Proof of  Theorem[\ref{tj}] is straightforward. Firstly, if $\sigma M$ is symmetric then Eq.[\ref{newIdentity0}]
devolves to  Eq.[\ref{uid}].
What if $\sigma M$ has no symmetry?
Eqs.[\ref{newIdentity0},\ref{newIdentity}] are linear in $M$.
Expand $M$ in terms of a linear combination of the 64 basis $8 \times 8$ matrices comprised of the 36 = 35 + 1 basis matrices
$M_s\,\in S_{8\times8}$ such that  $\sigma M_s$ is symmetric,
plus the 28 = 7 + 21 basis matrices
$M_a\,\in A_{8\times8}$ such that  $\sigma M_a$ is anti-symmetric, and verify the theorem component by component.
The set of 35 + 1 matrices $S_{8\times8}$ is given by
\begin{equation}
\label{symmetricM}
S_{8\times8} =
\left\{
\left.
{\tau}^{ (A) }{\tau}^{ (B) }{\tau}^{ (C) }
\right]
_
{
\{A,B,C\}
\in
\{1,\ldots,7\}
\&\&
A>B>C
}
, \,\mathbb{I}_{8\times8}
\right\}
,
\end{equation}
and each element of this set clearly verifies   Theorem[\ref{tj}].

The 7 + 21 matrices $M_a\,\in A_{8\times8}$ such that  $\sigma\,M_a$ is anti-symmetric are given by

\begin{equation}
\label{asymmetricM}
A_{8\times8} =
\left\{
\left.
{\tau}^{ (A) }
\right]_{A\in\{1,\ldots,7\}}
, \,
\left.
{\tau}^{ (A) }{\tau}^{ (B) }
\right]
_
{
\{A,B\}
\in
\{1,\ldots,7\}
\&\&
A>B
}
\right\}
.
\end{equation}

Each of
$
M_a \in \left\{
\left.
{\tau}^{ (A) }
\right]_{A\in\{1,\ldots,7\}}
\right\}
$
satisfies
$\tau _{ (\mu) }\,M_a\, \overline{\tau}^{ (\mu) }
\,=\,-4 M_a$
as well as
$\tau _{ (\mu) } \, \textrm{tr}\left(\overline{\tau}^{ (\mu) } \, M_a\right)=\,+8 M_a$.
Each of
$
M_a \in \left\{
\,
\left.
{\tau}^{ (A) }{\tau}^{ (B) }
\right]
_
{
\{A,B\}
\in
\{1,\ldots,7\}
\&\&
A>B
}
\right\}
$
satisfies
$\tau _{ (\mu) }\,M_a\, \overline{\tau}^{ (\mu) }
\,=\,+4 M_a$
as well as
$\tau _{ (\mu) } \, \textrm{tr}\left(\overline{\tau}^{ (\mu) } \, M_a\right)=\,0$.
Therefore each element of $A_{8\times8}$
satisfies Eqs.[\ref{newIdentity0},\ref{newIdentity}]
and the Theorem[\ref{tj}] is proven.
$\,\blacksquare$

\section{Bilocal Tetrad}
\label{Application2}

Let
$\mathbf{u} = \mathbf{u}(x^{ \alpha })  \in \, S^{(1)}(\mathbb{E}_{4,4})$ be a real eight component type-1 spinor field
(a section of the type-1 spinor bundle $S^{(1)}(\mathbb{E}_{4,4})$).
$\mathbf{u}$ is called the ``unit field" for reasons that are explained in  Section [\ref{sec:Algebra-signifi-spinor}].
In a quantum theory the $u^a$ satisfy commutation relations rather than anti-commutation relations
because of triality.
We  assume that
$
< \mathbf{u}, \mathbf{u} > = \widetilde{ {\mathbf{u}} } \; \sigma  \;  \mathbf{u} > 0
$
everywhere on  $\mathbb{E}_{4,4} = \pi\left(S^{(1)}(\mathbb{E}_{4,4})\right)$.

For brevity  a vielbein set of 8 independent vector fields is simply referred to as a tetrad (vierbein).
In this Section and the next we replace the indices $A, B, \ldots =  1, \ldots, 8$
with the indices $(\mu), (\nu), \ldots$,
where $\mu, \nu, \ldots =  1, \ldots, 8$,
in order to display this information is a more conventional form.
Summarizing,
$\alpha, \beta, \ldots, \mu, \nu, \ldots , a, b, \ldots \,=  1, \ldots, 8$.
We also employ
$\alpha_4, \beta_4, \ldots, \mu_4, \nu_4, \ldots   \,=  1, \ldots, 4$.

Let
$\psi \in S^{(2)}(\mathbb{E}_{4,4})$ denote a real eight component type-2
spinor field that realizes the bilocal  Cartesian  coordinates
$\mathbf{x} \,\equiv \,\left( x^{\alpha_4}, x^{4+\alpha_4} \right)$ of
$\pi\left(S^{(2)}(\mathbb{E}_{4,4})\right)$ $\cong$ $M_{3,1} \times {}^{*}\!M_{3,1}$.
The Cartesian coordinates
$x^{\alpha} \,=\, \left\{\mathbf{x}\right\}^{\alpha} $ are assumed to be $C^{\infty}$ functions of $\psi$,
$x^{\alpha} = x^{\alpha} (\psi)$
such that
$det( \frac{\partial x^{\alpha}}{\partial \psi^a}  ) \neq 0$, so that the
inverse
$\psi^a = \psi^a (x^{\alpha})$ always exists.
We abuse notation and write
$\mathbf{u} = \mathbf{u}(x^{ \alpha }) = \mathbf{u}(x^{ \alpha }(\psi)) = \mathbf{u}(\psi^a)$.
The mass dimension of $\psi$, $\left[\psi\right]$, is -1:
$\left[\psi\right]$ = LENGTH = 1/MASS = $[G^{1/2}]$ = [Planck length].

We define a spacetime tetrad $E^{(\mu)}$ with components $E^{(\mu)}_{\alpha}$ as
\begin{equation}
\label{vierbein}
E^{(\mu)}_{\alpha}  =  \frac{1}{\sqrt{\widetilde{\mathbf{u}} \, \sigma \, \mathbf{u}} }\; \widetilde{\mathbf{u}} \, \sigma \, \overline{\tau}^{(\mu)}\, \frac{\partial}{\partial x^{\alpha} }\psi.
\end{equation}

\emph{Remark}: Let $f : \mathbb{E}_{4,4}\rightarrow \mathbb{R}$ and
$
\frac{\partial \, f}{\partial x^{\alpha} }\,=\,f,_{\alpha}
$.
Let $r : \mathbb{E}_{4,4}\rightarrow \mathbb{R}^{+}$.
The tetrad $E^{(\mu)}$ may be made to transform
covariantly under the local projective transformation
%
\begin{eqnarray}
\label{lpt}
\mathbf{u} &\mapsto&  \mathbf{u'} \, = r(x^{\alpha}) \, \mathbf{u}
\nonumber \\
\psi &\mapsto&  {\psi'} \, = r(x^{\alpha}) \, \psi
\end{eqnarray}
by replacing the gradient operator
$\frac{\partial}{\partial x^{\alpha} }$ with
\begin{equation}
\label{eq:DA}
D_{\alpha}
=
\,  \mathbb{I}_{8 \times 8} \, \frac{\partial}{\partial x^{\alpha} }
\,-\,
\frac{1}{ {\widetilde{\mathbf{u}} \, \sigma \, \mathbf{u}}}
\tau_{(\mu)} \,
\frac{\partial\,\mathbf{u}}{\partial x^{\alpha} }
\,\otimes\,\widetilde{\mathbf{u}}\, \sigma \, \overline{\tau}^{(\mu)}
\end{equation}
because
$
D'_{\alpha} \,{\psi'} \, = \,
\left( r\,\psi,_{\alpha} + \,r,_{\alpha}\,\psi \right)
$
$
\,-\,
\frac{1}{ r^2\,{\widetilde{\mathbf{u}} \, \sigma \, \mathbf{u}}}
\tau_{(\mu)} \,
\left( r\,\mathbf{u},_{\alpha} + \,r,_{\alpha}\,\mathbf{u} \right)
\,\otimes\,r\,\widetilde{\mathbf{u}}\, \sigma \, \overline{\tau}^{(\mu)}\;
\left( r \, \psi\right)
$
$
= r \, D_{\alpha} \,{\psi}
$,
since
$
\frac{1}{ r^2\,{\widetilde{\mathbf{u}} \, \sigma \, \mathbf{u}}}
\tau_{(\mu)} \,
\left(  \,r,_{\alpha}\,\mathbf{u} \right)
\,\otimes\,r\,\widetilde{\mathbf{u}}\, \sigma \, \overline{\tau}^{(\mu)}\;
\left( r \, \psi\right)
=
\,r,_{\alpha}\,\psi \,
\left(  \frac{1}{ {\widetilde{\mathbf{u}} \, \sigma \, \mathbf{u}}}
\tau_{(\mu)} \,\mathbf{u}
\,\otimes\,\widetilde{\mathbf{u}}\, \sigma \, \overline{\tau}^{(\mu)}\right)
=
\,r,_{\alpha}\,\psi
$,
using
Eq.[\ref{uid}] or Eq.[\ref{newIdentity0}].

Therefore, if
\begin{equation}
\label{Er}
E^{(\mu)}_{\alpha}  =  \frac{1}{\sqrt{\widetilde{\mathbf{u}} \, \sigma \, \mathbf{u}} }\; \widetilde{\mathbf{u}} \, \sigma \, \overline{\tau}^{(\mu)}\, D_{\alpha}\,\psi.
\end{equation}

then
\begin{equation}
\label{rE}
E^{(\mu)}_{\alpha} \,\mapsto \, E'^{(\mu)}_{\alpha}   =   r(\mathbf{x}) \, E^{(\mu)}_{\alpha}
\end{equation}
under the local projective transformation Eq.[\ref{lpt}].

This local projective transformation generates a conformal transformation of the metric tensor.
$\blacksquare$

\begin{lemma}
\label{itj}
The inverse of the tetrad has components $E^{\alpha}_{(\mu)}$
\begin{equation}
\label{ivierbein}
E^{\alpha}_{(\mu)}  =   \frac{1}{\sqrt{\widetilde{\mathbf{u}} \, \sigma \, \mathbf{u}} }\; \frac{\partial x^{\alpha}}{\partial \psi} \tau_{(\mu)} \, \mathbf{u}.
\end{equation}

\emph{Proof}:

\begin{eqnarray}
\label{iEE}
E^{\alpha}_{(\mu)} \, E^{(\mu)}_{\beta}  &=&
\left(\frac{1}{\sqrt{\widetilde{\mathbf{u}} \, \sigma \, \mathbf{u}} }\; \frac{\partial x^{\alpha}}{\partial \psi} \tau_{(\mu)} \, \mathbf{u}\right)
\left( \frac{1}{\sqrt{\widetilde{\mathbf{u}} \, \sigma \, \mathbf{u}} }\; \widetilde{\mathbf{u}} \, \sigma \, \overline{\tau}^{(\mu)}\, \frac{\partial}{\partial x^{\beta} }\psi\right)
\nonumber \\
&=& \frac{1}{ {\widetilde{\mathbf{u}} \, \sigma \, \mathbf{u}} }\;
\frac{\partial x^{\alpha}}{\partial \psi}\,
\left( {\tau}_{(\mu)}\, \mathbf{u}\;\widetilde{\mathbf{u}} \, \sigma \, \overline{\tau}^{(\mu)}\right)\,
\frac{\partial\,\psi}{\partial x^{\beta} }
\nonumber \\
&=&
\frac{\partial x^{\alpha}}{\partial \psi}\,
\frac{\partial\,\psi}{\partial x^{\beta} }\;\;
\textrm{ by  Eq.[\ref{uid}] or Eq.[\ref{newIdentity0}]}
\nonumber \\
&=& \frac{\partial \,  x^{\alpha} }{\partial x^{\beta} }
\nonumber \\
&=&
\delta_{\beta}^{\alpha}
\end{eqnarray}

\emph{QED} $\,\blacksquare$

\end{lemma}

Since a matrix commutes with its inverse we also have
\begin{equation}
\label{LagrangianU1a}
E^{(\mu)}_{\alpha} \, E^{\alpha}_{(\nu)}  =   \delta_{(\nu)}^{(\mu)}.
\end{equation}

Let's look at an example.
We make the self-consistent assumption that there exists
a constant spacetime tetrad ${}^{(0)}\!E^{(\mu)}$ with constant components ${}^{(0)}\!E^{(\mu)}_{\alpha}$,
which might verify
${}^{(0)}\!E^{(\mu)}_{\alpha}\,=\,\delta^{(\mu)}_{\alpha}$,
for example.
Pick a constant unit field
$\mathbf{u}$ that satisfies
$\widetilde{\mathbf{u}} \, \sigma \, \mathbf{u} \, > \, 0$,
define
\begin{equation}
\label{psiupsi}
\psi =
\frac{1}{\sqrt{\widetilde{\mathbf{u}} \, \sigma \,\mathbf{u}} }\;
\tau_{(\nu)} \, \mathbf{u}  \; \;  {}^{(0)}\!E^{(\nu)}_{\beta} \, x^{\beta}
\end{equation}
(compare with the twistor \cite{Penrose:1988} type)
and compute
\begin{eqnarray}
\label{vierbein0}
{}^{(0)}\!E^{(\mu)}_{\alpha}  &=& \,
\frac{1}{\sqrt{\widetilde{\mathbf{u}} \, \sigma \,\mathbf{u}} }\;
\widetilde{\mathbf{u}} \, \sigma \, \overline{\tau}^{(\mu)}\, \frac{\partial}{\partial x^{\alpha} }\psi
\nonumber \\
&=&
\frac{1}{{\widetilde{\mathbf{u}} \, \sigma \,\mathbf{u}} }\;
\widetilde{\mathbf{u}} \, \sigma \, \overline{\tau}^{(\mu)}\,
\frac{\partial}{\partial x^{\alpha} }
\left(\tau_{(\nu)} \; \mathbf{u}  \;\; {}^{(0)}\!E^{(\nu)}_{\beta} \, x^{\beta}
\right)
\nonumber \\
&=&
\frac{1}{{\widetilde{\mathbf{u}} \, \sigma \,\mathbf{u}} }\;
\left(\widetilde{\mathbf{u}} \, \sigma \, \overline{\tau}^{(\mu)}\,
\tau_{(\nu)} \; \mathbf{u} \right) \;{}^{(0)}\!E^{(\nu)}_{\alpha}
\nonumber \\
&=&
{}^{(0)}\!E_{\alpha}^{{(\mu)}}\;\;\textrm{using the identity of Eq.[\ref{tautauID}]}
.
\end{eqnarray}


The pseudo-Riemannian metric associated to this tetrad field is
\begin{eqnarray}
\label{metric}
g_{\alpha \beta} &=& E^{(\mu)}_{\alpha} \; \eta_{{(\mu)} {(\nu)}}  \; E^{(\nu)}_{\beta}
\nonumber \\
&=& \frac{1}{ {\widetilde{\mathbf{u}} \, \sigma \, \mathbf{u}} }\;
\widetilde{\mathbf{u}} \, \sigma \, \overline{\tau}^{(\mu)}\, \frac{\partial}{\partial x^{\alpha} }\psi
 \; \eta_{{(\mu)} {(\nu)}} \;
\widetilde{\mathbf{u}} \, \sigma \, \overline{\tau}^{(\nu)}\, \frac{\partial}{\partial x^{\beta} }\psi
\nonumber \\
&=& \frac{1}{ {\widetilde{\mathbf{u}} \, \sigma \, \mathbf{u}} }\;
\left(\widetilde{\mathbf{u}} \, \sigma \, \overline{\tau}^{(\mu)}\, \frac{\partial}{\partial x^{\alpha} }\psi\right)^{T}
 \; \eta_{{(\mu)} {(\nu)}} \;
\widetilde{\mathbf{u}} \, \sigma \, \overline{\tau}^{(\nu)}\, \frac{\partial}{\partial x^{\beta} }\psi
\nonumber \\
&=& \frac{1}{ {\widetilde{\mathbf{u}} \, \sigma \, \mathbf{u}} }\;
\frac{\partial}{\partial x^{\alpha} }\widetilde{\psi}\, \sigma \, {\tau}^{(\mu)}\, \mathbf{u}
 \; \eta_{{(\mu)} {(\nu)}} \;
\widetilde{\mathbf{u}} \, \sigma \, \overline{\tau}^{(\nu)}\, \frac{\partial}{\partial x^{\beta} }\psi
\,
\textrm{ ,  by  Eq.[\ref{tau-bar}]}
\nonumber \\
&=& \frac{1}{ {\widetilde{\mathbf{u}} \, \sigma \, \mathbf{u}} }\;
\frac{\partial}{\partial x^{\alpha} }\widetilde{\psi}\, \sigma \,\left( {\tau}_{(\nu)}\, \mathbf{u}
\;
\widetilde{\mathbf{u}} \, \sigma \, \overline{\tau}^{(\nu)}\right)\, \frac{\partial}{\partial x^{\beta} }\psi
\nonumber \\
&=& \frac{\partial\,\widetilde{\psi}}{\partial x^{\alpha} }\, \sigma \, \frac{\partial\,\psi}{\partial x^{\beta} }\;\;
\textrm{ using  Eq.[\ref{uid}] or Eq.[\ref{newIdentity0}]}
\nonumber \\
&=& \frac{\partial \, {\psi}^{a}}{\partial x^{\alpha} }\, \sigma_{a b} \, \frac{\partial \, \psi^{b}}{\partial x^{\beta} }
\nonumber \\
&=& \frac{\partial \, {\psi}^{a}}{\partial x^{\alpha} } \, \frac{\partial \, \psi^{b}}{\partial x^{\beta} }\, \sigma_{a b},
\end{eqnarray}
which is not an induced metric but, as one may expect,
is the coordinate-transform of $\sigma_{a b}$.
The 4 + 4 = 8 dimensional spacetime $\pi\left( S^{(1)}(\mathbb{E}_{4,4}) \right)$
endowed with this metric
has zero curvature.

\section{Schwinger real representation of QED}
\label{Application1}

Julian Schwinger \cite{Schwinger64} has given a representation of charged fermion field operators for an electron
in terms of
real anti-commuting 8-component spinor fields.
Therefore it may be of interest to evaluate,  using the new identity Eq.[\ref{newIdentity0}]
and the above tetrad,
and with arbitrary  Schwinger spinor (bilocal) fields $F$ and $H$, the operator
\begin{eqnarray}
\label{td}
H\!&&\!\! \gamma^{(\mu)}\;\;  E^{\alpha}_{(\mu)}\;\;\frac{\partial}{\partial x^{\alpha} } F\;=\;
\,H\, \overline{\tau}^{(\mu)}\,
\frac{1}{\sqrt{\widetilde{\mathbf{u}} \, \sigma \, \mathbf{u}} }\; \frac{\partial x^{\alpha}}{\partial \psi} \tau_{(\mu)} \, \mathbf{u}
\; \frac{\partial}{\partial x^{\alpha} } F
\nonumber \\
&=&
\frac{1}{\sqrt{\widetilde{\mathbf{u}} \, \sigma \, \mathbf{u}} }\;
H\,\overline{\tau}^{(\mu)}\,
\frac{\partial F}{\partial \psi} \tau_{(\mu)} \, \mathbf{u}
\nonumber \\
&=&
\left[
\,-\,\left(\tau _{ (\mu) }\right)_{\phantom{c }d}^a \left(\overline{\tau}^{ (\mu) }\right)_{\phantom{a }b}^c
\,+\,
2 \left(\delta _b^a \delta _d^c+\delta _d^a \delta _b^c-\sigma ^{a c} \sigma _{b d}\right)
\right]
H_c\,
\frac{u^b}{\sqrt{\widetilde{\mathbf{u}} \, \sigma \, \mathbf{u}} }
\,\frac{\partial F^d}{\partial \psi^a},
\nonumber \\
\end{eqnarray}
which may easily be further reduced.

\section{Algebraic significance of the spinor $\mathbf{u}$, the unit field}
\label{sec:Algebra-signifi-spinor}

Let $\mathbf{u} \in S^{(1)}(\mathbb{E}_{4,4})$ be a  type-1 spinor field (a section of the type-1 spinor bundle $S^{(1)}(\mathbb{E}_{4,4})$), with
$
< \mathbf{u}, \mathbf{u} > = \widetilde{ {\mathbf{u}} } \; \sigma  \;  \mathbf{u} > \,0
$
everywhere on the base space $\mathbb{E}_{4,4}$, but being otherwise arbitrary.
$\mathbf{u}$, may be called a ``unit field".
One may define a special $\mathbb{E}_{4,4}$
frame field  $\mathfrak{F}$ in terms of $\mathbf{u}$ and the tau matrices as follows. Let M be the real $8 \times 8$
matrix defined by
\begin{eqnarray}
M &=&  \frac{1}{{\widetilde{\mathbf{u}} \, \sigma \, \mathbf{u}} }\; \mathbf{u}\,\otimes \widetilde{\mathbf{u}}\,\sigma =
\frac{1}{{\widetilde{\mathbf{u}} \, \sigma \, \mathbf{u}} }\;\mathbf{u}\, \widetilde{\mathbf{u}}\,\sigma
\nonumber \\
M^a_{\phantom{a} b} &=&  \frac{1}{{\widetilde{\mathbf{u}} \, \sigma \, \mathbf{u}} }\; u^a\,  u^c \, \sigma_{c b}
\label{V}
\end{eqnarray}
Then M obeys Eq.[\ref{sU}] and transforms under $\overline{S0(4,4; \mathbb{R})}$ according to
Eq.[\ref{U}].
Using Eq.[\ref{uid}] or Eq.[\ref{newIdentity0}] to evaluate
$
\tau_A \; M \; \overline{\tau}^A
$
yields
\begin{equation}
\mathbb{I}_{8 \times 8}  = \frac{1}{{\widetilde{\mathbf{u}} \, \sigma \, \mathbf{u}} }\;
\tau_A \;\left( \mathbf{u}\, \widetilde{\mathbf{u}}\,\sigma \right) \; \overline{\tau}^A
=
\left(\frac{1}{\sqrt{\widetilde{\mathbf{u}} \, \sigma \, \mathbf{u}} }\;\tau_A \; \mathbf{u}\right)\,
\left(\frac{1}{\sqrt{\widetilde{\mathbf{u}} \, \sigma \, \mathbf{u}} }\;\widetilde{\mathbf{u}}\sigma \; \overline{\tau}^A\right)
\label{uuid}
\end{equation}
This is a resolution of the identity on $\mathbb{E}_{4,4}$. Alternatively this relation may be
interpreted as a completeness condition verified by the $\mathbb{E}_{4,4}$
orthogonal frame $\mathfrak{F}$
whose components  $\mathfrak{F}^{a}_{A}$ are given by
\begin{equation}
\mathfrak{F}^{a}_{A} = \frac{1}{\sqrt{\widetilde{\mathbf{u}} \, \sigma \, \mathbf{u}} }\; \tau^{\phantom{A}\,a}_{A\phantom{\,a}\,b} \, \mathbf{u}^b
\label{F}
\end{equation}
and its inverse is
\begin{equation}
\mathfrak{F}^A_a =\frac{1}{\sqrt{\widetilde{\mathbf{u}} \, \sigma \, \mathbf{u}} }\;  \mathbf{u}^c \sigma_{c b} \overline{\tau}^{A\,b}_{\phantom{A\,b}a}
\label{iF}
\end{equation}

Accordingly Eq.[\ref{uuid}] may be expressed in index notation as
\begin{equation}
\{  \mathbb{I}_{8 \times 8}  \}^a_b =
\delta^a_b = \mathfrak{F}^{a}_{A} \,\mathfrak{F}^A_b
\label{FiF}
\end{equation}
Since a matrix commutes with its inverse we also have
\begin{equation}
\delta^A_B = \mathfrak{F}^A_a\,\mathfrak{F}^{a}_{B}.
\label{iFF}
\end{equation}

We have defined an oriented spinor basis $e_a$ of  $S_x^{(1)}(\mathbb{E}_{4,4})$ in Eq[\ref{neaeb}]
and  an oriented vector basis $\epsilon_A$
of $T_x(\mathbb{E}_{4,4})$ in Eq[\ref{neAeB}].
The two are related by
\begin{eqnarray}
\epsilon_A &=& e_a \, \mathfrak{F}^{a}_{A}\;\;\textrm{  and}
\nonumber \\
e_a \, &=&  \, \epsilon_A \mathfrak{F}_{a}^{A}
\label{eAea}
\end{eqnarray}

\subsection{Split octonion algebra over $\mathbb{R}$, $\mathbb{O}_{s}(\mathbb{R})$}
Let  $\mathbb{O}_{s}(\mathbb{R})$ denote the split octonion algebra over $\mathbb{R}\;$
\cite{Zorn:1931},
\cite{Springer:582165},
\cite{Harevy1990:9780123296504},
\cite{NASH90},
\cite{Nash2010},
\cite{Nash2012}
.

A nonassociative alternative multiplication of the oriented spinor basis
$e_a$ (respectively, oriented vector basis $\epsilon_A$) may be defined \cite{NASH90} that endows
the real vector space $S_x^{(1)}(\mathbb{E}_{4,4})$ (respectively, $T_x(\mathbb{E}_{4,4})$) with the structure of a
normed nonassociative
algebra with multiplicative unit that is isomorphic to the
split octonion algebra over $\mathbb{R}$, $\mathbb{O}_{s}(\mathbb{R})$.
This is accomplished by specifying the
multiplication constants $m^c_{a b}$ (respectively, $m^C_{A B}$) of the algebra, which verify
\begin{eqnarray}
e_a \, e_b  &=& e_c \;m^c_{a b}
\nonumber \\
\epsilon_A \, \epsilon_B  &=& \epsilon_C \;m^C_{A B}
\label{eaeb}
\end{eqnarray}
The set of multiplication constants $m^c_{a b}$ (respectively, $m^C_{A B}$) is defined by  \cite{NASH90}
\begin{eqnarray}
m^c_{a b} &=& \mathfrak{F}^A_a\, \tau^{\phantom{A}\,c}_{A\phantom{\,c}\,b}
\nonumber \\
m^C_{A B}&=& \mathfrak{F}^C_c\, \tau^{\phantom{A}\,c}_{A\phantom{\,c}\,b}\,\mathfrak{F}^b_B.
\label{mcab}
\end{eqnarray}
It has been shown that the  nonassociative product defined by Eq.[\ref{eaeb}] (respectively,  Eq.[\ref{mcab}])
of the spinor basis $e_a$ (respectively, of the vector basis $\epsilon_A$ )
endows the respective  real vector space
with the
structure of the split octonion algebra over the reals \cite{NASH90}.
This is explicit in the  multiplication table below, which employs the representation of
the tau matrices given in  Appendix 3
and
$u^a \, = \, \frac{1}{\sqrt{2}}(0,1,0,0,0,1,0,0)$,
which is an eigenvector of $\sigma$ with eigenvalue +1.

\subsection{Multiplicative identity}

An element $\Psi \in \mathbb{O}_{s}(\mathbb{R})$ may be realized as
\begin{eqnarray}
\Psi = e_a  \,\psi^a &=& \epsilon_A \, \hat{\psi}^A
\nonumber \\
\hat{\psi}^A = \mathfrak{F}^A_a\, \psi^a\,
&\Leftrightarrow&\,
\psi^a = \mathfrak{F}^a_A\, \hat{\psi}^A\,.
\label{Psi}
\end{eqnarray}

The normalized fiducial  unit field
$\mathbb{O}_{s}(\mathbb{R})\, \ni\, \frac{1}{\sqrt{< \mathbf{u}, \mathbf{u} >} }\; \mathbf{u}$
$= \,\frac{1}{\sqrt{\widetilde{\mathbf{u}} \, \sigma \, \mathbf{u}} }\;e_a \, u^a $ =
$\frac{1}{\sqrt{\widetilde{\mathbf{u}} \, \sigma \, \mathbf{u}} }\; \tau^{\phantom{A}\,a}_{8\phantom{\,a}\,b} \, u^b$
$= e_a \, \mathfrak{F}^{a}_{8} = \epsilon_8$
=
\textbf{\emph{multiplicative identity}} element
of the
split octonion algebra  $\mathbb{O}_{s}(\mathbb{R})$ \cite{Nash2010}:
\begin{equation}
\textrm{Multiplicative identity} =
\frac{1}{\sqrt{\widetilde{\mathbf{u}} \, \sigma \, \mathbf{u}} }\;
e_a\, u^a = \frac{1}{\sqrt{< \mathbf{u}, \mathbf{u} >} }\;  \mathbf{u}
= \epsilon_8.
\label{unit}
\end{equation}


$\textrm{Multiplication Table }\\\epsilon_A \times \epsilon_B =$
\begin{displaymath}
\begin{array}{|c||c|c|c|c|c|c|c|c|}
\hline
\textrm{\backslashbox{$\epsilon_A =$}{$\epsilon_B =$}}  &\;
 \epsilon_{1} &\;  \epsilon_{2} &\;  \epsilon_{3} &\;  \epsilon_{4} &\;  \epsilon_{5} &\;  \epsilon_{6} &\;   \epsilon_{7} &\; \epsilon_{8}\; \\
   \hline\hline
 \epsilon_{1} & -\epsilon_{8} & \epsilon_{3} & -\epsilon_{2} & -\epsilon_{5} & \epsilon_{4} & -\epsilon_{7} & \epsilon_{6} & \epsilon_{1} \\
\hline
 \epsilon_{2} & -\epsilon_{3} & -\epsilon_{8} & \epsilon_{1} & -\epsilon_{6} & \epsilon_{7} & \epsilon_{4} & -\epsilon_{5} & \epsilon_{2} \\
\hline
 \epsilon_{3} & \epsilon_{2} & -\epsilon_{1} & -\epsilon_{8} & -\epsilon_{7} & -\epsilon_{6} & \epsilon_{5} & \epsilon_{4} & \epsilon_{3} \\
\hline
 \epsilon_{4} & \epsilon_{5} & \epsilon_{6} & \epsilon_{7} & \epsilon_{8} & \epsilon_{1} & \epsilon_{2} & \epsilon_{3} & \epsilon_{4} \\
\hline
 \epsilon_{5} & -\epsilon_{4} & -\epsilon_{7} & \epsilon_{6} & -\epsilon_{1} & \epsilon_{8} & \epsilon_{3} & -\epsilon_{2} & \epsilon_{5} \\
\hline
 \epsilon_{6} & \epsilon_{7} & -\epsilon_{4} & -\epsilon_{5} & -\epsilon_{2} & -\epsilon_{3} & \epsilon_{8} & \epsilon_{1} & \epsilon_{6} \\
\hline
 \epsilon_{7} & -\epsilon_{6} & \epsilon_{5} & -\epsilon_{4} & -\epsilon_{3} & \epsilon_{2} & -\epsilon_{1} & \epsilon_{8} & \epsilon_{7} \\
\hline
 \epsilon_{8} & \epsilon_{1} & \epsilon_{2} & \epsilon_{3} & \epsilon_{4} & \epsilon_{5} &   \epsilon_{6} & \epsilon_{7} & \epsilon_{8}\\ \hline
\end{array}
\label{mus}
\end{displaymath}


\section{Concluding remark}

The reduced generators (i.e., the Dirac matrices) of the pseudo Clifford algebra $\mathbb{C}_{4,4}$
defined on $\mathbb{E}_{4,4}$ satisfy a remarkable identity Eq.[\ref{newIdentity}]
that
defines invertible linear mappings  between each of the two
types of $\overline{S0(4,4; \mathbb{R})}$ spinor and the ${S0(4,4; \mathbb{R})}$ vector, thereby
admitting   matrix representations of triality on this spacetime $\mathbb{E}_{4,4}$.
The trialities are given below in Eqs[\ref{QA}] and [\ref{QAB}].

\section{Appendix 1: Transformation under action of $\overline{S0(4,4; \mathbb{R})}$}


The special Lorentz transformation properties of the theory
may be determined by constructing a real reducible $16 \times 16$ matrix representation
of $\overline{S0(4,4; \mathbb{R})}$ utilizing the  irreducible generators
$t^A$ , \; A = 1, ... , 8 of the (pseudo-) Clifford algebra $C_{4,4}$.
Following Lord's general procedure \cite{Lord68} we define
the  irreducible generators
$t^A$ as
\begin{eqnarray}
\label{tA}
t^A
&=&
\left(
\begin{array}{cc}
 0 & \overline{\tau}^A  \\
 \tau^A &  0
\end{array}
\right).
\end{eqnarray}

Let $g \in \overline{S0(4,4; \mathbb{R})}$.
The  $16 \times 16$ basic spinor representation of ${S0(4,4; \mathbb{R})}$
is reducible into the two real $8 \times 8$ inequivalent irreducible
spinor representations ${D_{(1)}}(g)$ and ${D_{(2)}}(g)$ of $\overline{S0(4,4; \mathbb{R})}$.
The reduced generators of the two real $8 \times 8$
spinor representations ${D_{(1)}}(g)$ and ${D_{(2)}}(g)$ of
$\overline{S0(4,4; \mathbb{R})}$  follow
from the calculation of the  infinitesimal generators
\begin{eqnarray}
t^A t^B - t^B t^A
&=&
\left(
\begin{array}{cc}
  \overline{\tau}^A \tau^B -   \overline{\tau}^B \tau^A & 0\\
0 &  { \phantom{=}  \tau^A  \overline{\tau}^B - \tau^B \overline{\tau}^A \phantom{=} }
\end{array}
\right)\nonumber \\
&=&
4
\left(
\begin{array}{cc}
{{D_{(1)}}}{}^{A B} & 0\\
0 &  {{D_{(2)}}}{}^{A B}
\end{array}
\right),
\label{commutat}
\end{eqnarray}
of the 16-component
spinor representation of  $\overline{S0(4,4; \mathbb{R})}$.
We see, as is in fact well
known from the general theory, that  the 16-component
spinor representation of  $\overline{S0(4,4; \mathbb{R})}$ is the direct sum of
two (inequivalent)
real $8 \times 8$ irreducible  spinor representations  ${D_{(1)}} = {D_{(1)}}(g)$ and ${D_{(2)}} = {D_{(2)}}(g)$ of
$\overline{S0(4,4; \mathbb{R})} \ni g$ that are generated by ${{D_{(1)}}}{}^{A B} \textrm{ and }
 {{D_{(2)}}}{}^{A B}$  respectively, where
\begin{equation}
4 \, {{D_{(1)}}}{}^{A B} =
\overline{\tau}^A \tau^B - \overline{\tau}^B \tau^A
\label{D1}
\end{equation}
and
\begin{equation}
4 \, {{D_{(2)}}}{}^{A B} =
\tau^A \overline{\tau}^B - \tau^B \overline{\tau}^A
\label{D2}
\end{equation}

For completeness we remark that the generators of
the two spinor types are images of the projection operators
\begin{eqnarray}
\label{t99}
\chi_{\pm} &=& \frac{1}{2}(1 \pm t^9)
\nonumber \\
\chi_{+} &=&
\left(
\begin{array}{cc}
 \mathbb{I}_{8 \times 8} & 0  \\
 0 &  0
\end{array}
\right)
\nonumber \\
\chi_{-} &=&
\left(
\begin{array}{cc}
 0 & 0  \\
0 &  \mathbb{I}_{8 \times 8}
\end{array}
\right) ,
\end{eqnarray}
where
\begin{equation}
t^{9} = t^{1}t^{2}t^{3}t^{4}t^{5}t^{6}t^{7}t^{8}
=
\left(
\begin{array}{cc}
 \overline{\tau}^0 & 0  \\
 0 &  \tau^0
\end{array}
\right) .
\label{t9}
\end{equation}
Here
\begin{equation}
\overline{\tau}^{0} = \overline{\tau}^{1}\tau^{2}\overline{\tau}^{3}\tau^{4}\overline{ \tau}^{5}\tau^{6}\overline{\tau}^{7}
= \tau^{1}\tau^{2}\tau^{3}\tau^{4}\tau^{5}\tau^{6}\tau^{7}
\label{tau7}
\end{equation}
and
\begin{equation}
\tau^{0} = \tau^{1}\; \overline{\tau}^{2} \; \tau^{3}\;
\overline{\tau}^{4}\;\tau^{5}\; \overline{\tau }^{6} \; \tau^{7}
= -\tau^{1}\tau^{2}\tau^{3}\tau^{4}\tau^{5}\tau^{6}\tau^{7}
= -\overline{\tau}^{0}
\label{tau-7}
\end{equation}

The
representation of the tau matrices is irreducible.
$
\overline{\tau}^{0}
$
has square equal to $+ \mathbb{I}_{8 \times 8} $
and commutes with each of the $\tau^{A}$ matrices (and
therefore with all of their products).
Therefore we conclude that
$\overline{\tau}^{0} = \pm \mathbb{I}_{8 \times 8}$ in any irreducible representation.

Let
$
\omega_{A B} = - \omega_{B A} \in \mathbb{R}
$,
$A, B = 1, \ldots,8$,
enumerate a set of 28  real parameters that coordinatize
$g = g(\omega) \in \overline{S0(4,4; \mathbb{R})}$.
Also,
let
$L = L(g) \in SO(4, 4; \mathbb{R})$ have matrix elements $L^{A}_{\phantom{A} B} $,
$\omega^{\sharp}$
denote the real $8 \times 8$ matrix with matrix elements
$
\omega^A_{\phantom{A} B} = \mathbb{G}^{A C} \omega_{C B}
$,
$
\omega_{1} = \frac{1}{2}\omega_{A B} {D_{(1)}}^{A B}
$
and
$
\omega_{2} = \frac{1}{2}\omega_{A B} {D_{(2)}}^{A B}
$.
We find that
\begin{eqnarray}
\label{reps}
{D_{(1)}} &=& {D_{(1)}}(g)
= \exp{\left(\frac{1}{2}\omega_{1}\right)}
\nonumber \\
{D_{(2)}} &=& {D_{(2)}}(g)
= \exp{\left(\frac{1}{2}\omega_{2}\right)}
\nonumber \\
L^{A}_{\phantom{A} B} &=& L^{A}_{\phantom{A} B}(g)
= {\left\{\exp{\left(\omega^{\sharp}\right)}\right\}}^{A}_{\phantom{A} B}
\end{eqnarray}
where, under the action of $\overline{S0(4,4; \mathbb{R})}$,
\begin{equation}
\widetilde{{D_{(1)}}}{}^{A B}\sigma = -\sigma {D_{(1)}}^{A B}
\Rightarrow
\widetilde{D}_{(1)}\sigma = \sigma {D_{(1)}}^{-1}
\label{d1}
\end{equation}
\begin{equation}
\widetilde{{D_{(2)}}}{}^{A B}\sigma = -\sigma {D_{(2)}}^{A B}
\Rightarrow
\widetilde{D}_{(2)}\sigma = \sigma {D_{(2)}}^{-1}
\label{d2}
\end{equation}
\begin{eqnarray}
L^{A}_{\phantom{A} C} \mathbb{G}_{A\,B} L^{B}_{\phantom{B} D} &=&
 \mathbb{G}_{C\,D}
=
\left\{
\widetilde{L} \mathbb{G} L
\right\}{}_{C\,D}
\label{Lt}
\end{eqnarray}
\begin{equation}
\label{Ltaubar}
L^{A}_{\phantom{A} B}\overline{\tau}^B = {{D_{(1)}}}^{-1} \, \overline{\tau}^A {{D_{(2)}}}
\end{equation}
\begin{equation}
\label{Ltau}
L^{A}_{\phantom{A} B}\tau^B = {{D_{(2)}}}^{-1} \, \tau^A {D_{(1)}}
\end{equation}
The canonical 2-1 homomorphism
$\overline{S0(4,4; \mathbb{R})} \rightarrow S0(4,4; \mathbb{R}): g \mapsto L(g)$ is given by
\begin{equation}
8 \, L^{A}_{\phantom{A} B} = tr \left( {{D_{(1)}}}^{-1} \, \overline{\tau}^A
{D_{(2)}} \, \tau^C\right) \mathbb{G}_{C B},
\label{L}
\end{equation}
where  tr denotes the trace.
Note that ${D_{(1)}(g(\omega))} = {D_{(2)}(g(\omega))}$ when
$
\omega_{A 8} = 0
$,
i.e., when one restricts $\overline{S0(4,4; \mathbb{R})}$ to
\begin{eqnarray}
\label{eqn:SO34}
&&
\overline{S0(3,4; \mathbb{R})}
=
\left\{
g \in \overline{S0(4,4; \mathbb{R})} \;
\left.\right|\right.
\nonumber \\
&g&=
\left.
\left(
\begin{array}{cc}
 \exp{\left(\frac{1}{4}  \omega_{A B} {D_{(1)}}^{A B} \right)} & 0\\
 0 &  \exp{\left(\frac{1}{4}  \omega_{A B} {D_{(2)}}^{A B} \right)}
\end{array}
\right)
\textrm{ and }
\omega_{A 8} = 0
\right\}
\end{eqnarray}
This is one of the real forms of Spin$(7,\mathbb{C})$.

\section{Appendix 2: Triality and $\overline{S0(4,4; \mathbb{R})}$ covariant multiplications}

Let $V_1, V_2$, and $V_3$ be vector spaces over $\mathbb{R}$.
A duality is a nondegenerate bilinear map
$V_1 \times V_2   \rightarrow \mathbb{R}$.
A triality is a nondegenerate trilinear map
$V_1 \times V_2  \times V_3  \rightarrow \mathbb{R}$.
A triality may be associated with a bilinear map
that some authors call a ``multiplication" \cite{Baez}
by dualizing,
$V_1 \times V_2   \rightarrow  {}^{*}{V_3} \cong V_3$.

Let $\mathbf{u}$ denote the unit field and
let
$\psi_{(1)} \in S_x^{(1)}(\mathbb{E}_{4,4})$
and
$\psi_{(2)} \in S_x^{(1)}(\mathbb{E}_{4,4})$
.
%
%
Under the action of $\overline{S0(4,4; \mathbb{R})}$
we assume that
$
\mathbf{u} \mapsto \overline{\mathbf{u}} = {D_{(1)}} \;  \mathbf{u}
$,
$
\psi_{(1)} \mapsto \overline{\psi}_1 = {D_{(1)}} \;  \psi_{(1)}
$
and
$
\psi_{(2)} \mapsto \overline{\psi}_2 = {D_{(2)}} \;  \psi_{(2)}.
$
Consider the following two multiplications that possess  covariant transformation laws
under the action of
$
\overline{S0(4,4; \mathbb{R})}
\Rightarrow
{S0(4,4; \mathbb{R})}
$.
The first  multiplication
$
{m_1}^{A}\;: \mathbb{E}_{4,4} \times \mathbb{E}_{4,4} \rightarrow \mathbb{E}_{4,4}
$
is defined by
\begin{equation}
\label{QA}
{Q}^{A} =
\frac{1}{\sqrt{\widetilde{\mathbf{u}} \, \sigma \, \mathbf{u}} }\;
\widetilde{ {\mathbf{u}} } \; \sigma  \;   \overline{\tau}^A \,
{\psi_{(2)}} .
\end{equation}
For fixed $u^a$, ${Q}^{A} \in \mathbb{E}_{4,4}$ depends on 8 real parameters
arranged into the type-2 spinor $\psi_{(2)}$.

The second multiplication
$
{m_2}^{A B}\;: \mathbb{E}_{4,4} \times \mathbb{E}_{4,4} \rightarrow V_3
$
has an image in %
$V_3 \cong \mathbb{E}_{4,4}\times\mathbb{E}_{4,4}$, and
depends on 8 real parameters (for fixed $u^a$)
arranged into the type-1 spinor $\psi_{(1)}$:
\begin{equation}
\label{QAB}
{Q}^{A B} =
\frac{1}{\sqrt{\widetilde{\mathbf{u}} \, \sigma \, \mathbf{u}} }\;
\widetilde{ {\mathbf{u}} } \; \sigma  \;   \overline{\tau}^A \,{\tau}^B \,
{\psi_{(1)}} .
\end{equation}
For fixed $\mathbf{u}$,
${Q}^{A B}$ possesses only 8 degrees of freedom corresponding to  the 8 independent degrees of freedom of ${\psi_{(1)}}$, so we also refer
to this map as a ``multiplication."

Eq.[\ref{QAB}] may be easily be solved for the components ${\psi_{(1)}}^a \, = \, {\psi_{(1)}}^a({Q}^{A B})$.
Consider
\begin{eqnarray}
\label{para}
\frac{1}{\sqrt{\widetilde{\mathbf{u}} \, \sigma \, \mathbf{u}} }\;
\left( \overline{\tau}_B  \right) \;  \left( {Q}^{A B} \; {\tau}_{(A)}\, \mathbf{u}\;\right)\, &=&
\frac{1}{ {\widetilde{\mathbf{u}} \, \sigma \, \mathbf{u}} }\;
\left( \overline{\tau}_B  \right) \;
\left( {\tau}_{(A)}\, \mathbf{u}\;\right)\,
\left(\,
\widetilde{ {\mathbf{u}} } \; \sigma  \;   \overline{\tau}^A \,{\tau}^B \,{\psi_{(1)}}
\right)
\nonumber \\
&=&
\frac{1}{ {\widetilde{\mathbf{u}} \, \sigma \, \mathbf{u}} }\;
\left( \overline{\tau}_B  \right) \;
\left( {\tau}_{(A)}\, \mathbf{u}\;\widetilde{\mathbf{u}} \, \sigma \, \overline{\tau}^{(A)}\right)\,
{\tau}^B \, {\psi_{(1)}}
\nonumber \\
&=&
\left( \overline{\tau}_B  \right) \;
{\tau}^B \,{\psi_{(1)}}
\;\;
\textrm{ by  Eq.[\ref{uid}] or Eq.[\ref{newIdentity0}]}
\nonumber \\
&=&
\left( \overline{\tau}_B \,{\tau}^B \right) \; {\psi_{(1)}}
\nonumber \\
&=&
8 \, {\psi_{(1)}}
\end{eqnarray}
Similarly,

\begin{equation}
\label{sQA}
{\psi_{(2)}} \, =\,
\frac{1}{\sqrt{\widetilde{\mathbf{u}} \, \sigma \, \mathbf{u}} }\;{\tau}_A \,\mathbf{u}\; {Q}^{A}.
\end{equation}

In   this paragraph Greek indices run from 1 to 4,
$\alpha, \beta, \ldots, \mu, \nu, \ldots \,=  1, \ldots, 4$,
while Latin  continue to run from 1 to 8,
$\;A, B, \ldots,  a, b, \ldots \,=  1, \ldots, 8$.
It is convenient to define a $SO(3,1; \mathbb{R})$-invariant symplectic structure  $\Omega$ on
$\mathbb{E}_{4,4}$
(and a complex structure on the split octonion algebra) by
\begin{eqnarray}
\Omega
&=&
\left(
\begin{array}{cc}
 0 & 1\\
 -1 &  0
\end{array}
\right)
\label{omega}
\end{eqnarray}
where 0 denotes the 4 x 4 zero matrix and 1 denotes the 4 x 4 unit matrix.
The ${Q}^{A B}$ may be represented in terms of an arbitrary antisymmetric $M_{3,1}$ rank 2 tensor
$\mathbb{F}^{\beta\,\alpha} = -\mathbb{F}^{\alpha\,\beta}$
and two  $S0(3,1; \mathbb{R})$ scalars $x_{4}$ and  $x_{8}$ according to
\begin{equation}
\label{QABF}
{Q}^{A B} =
{\left\{
\left(
\begin{array}{cc}
 \mathbb{F}^{\alpha\,\beta} & {}^{*}\mathbb{F}^{\alpha\,}_{\phantom{\alpha\,}\beta} \\
 {}^{*}\mathbb{F}_{\alpha\,}^{\phantom{\alpha\,}\beta} & \mathbb{F}_{\alpha\,\beta}
\end{array}
\right)
\right\}}^{A B}
+ {Q}^{4 8} \; {\Omega}^{A B}
+ {Q}^{8 8} \; {\mathbb{G}}^{A B}
,
\end{equation}
where ${}^{*}\mathbb{F}^{\alpha\,\beta}$ is dual to $\mathbb{F}^{\alpha\,\beta}$ and defined by
${}^{*}\mathbb{F}^{\mu \nu} = - \frac{1}{2} \epsilon^{\alpha \beta \mu \nu} \mathbb{F}_{\alpha\,\beta}$.
Note that
$
Q^{[A\,B]}
=
 \frac{1}{2}\left(Q^{A\,B} - Q^{B\,A} \right)
 $
is independent of ${Q}^{8 8}$.


Clearly, in order for Eq.[\ref{QABF}] to possess physical significance the action of
$\overline{S0(4,4; \mathbb{R})}$ must be restricted to $\overline{S0(3,1; \mathbb{R})}$
in a manner that links transformations of $x^5, x^6, x^7, x^8$ to $x^1, x^2, x^3, x^4$.

\subsection{Covariance of maps under $\overline{S0(4,4; \mathbb{R})}$}

Let
$
\mathbf{u} \mapsto \overline{\mathbf{u}} = {D_{(1)}} \;  \mathbf{u}
,\;
\psi_{(1)} \mapsto \overline{\psi}_1 = {D_{(1)}} \;  \psi_{(1)}
\;\textrm{ and }
\psi_{(2)} \mapsto \overline{\psi}_2 = {D_{(2)}} \;  \psi_{(2)}
$
under $\overline{S0(4,4; \mathbb{R})}$.
Consider the transformation law for the
${Q}^{A} \mapsto \overline{Q}^{A}$:\\
$
\overline{Q}^{A}
=
\widetilde{ \overline{\mathbf{u}} } \; \sigma  \; \overline{\tau}^A \, \overline{\psi}_2 \,
=
\widetilde{ {{D_{(1)}}} \;\mathbf{u} } \; \sigma  \; \overline{\tau}^A \,  {{D_{(2)}}} \; {\psi}
$

$
\phantom{ \overline{Q}^{A} }
=
\widetilde{ {\mathbf{u}} } \; \sigma  \;  {{D_{(1)}}}^{-1} \; \overline{\tau}^A \,  {{D_{(2)}}}
 {\psi}
$

$
\phantom{ \overline{Q}^{A} }
=
L^{A}_{\phantom{A} B}\;
\widetilde{ {\mathbf{u}} } \; \sigma  \;   \overline{\tau}^B \,
 {\psi}
$
$
\phantom{ \overline{Q}^{A} }
=
L^{A}_{\phantom{A} B}\;{Q}^{B}
$,\\
which follows from  Eq.[\ref{Ltaubar}],
Also
${Q}^{A B} \mapsto \overline{Q}^{A B}$:\\
$
\overline{Q}^{A B}
=
\widetilde{ \overline{\mathbf{u}} } \; \sigma  \; \overline{\tau}^A \,{\tau}^B \, \overline{\psi}_1 \,
=
\widetilde{ {{D_{(1)}}} \;\mathbf{u} } \; \sigma  \; \overline{\tau}^A \,{\tau}^B \,  {{D_{(1)}}} \; {\psi_{(1)}}
$

$
\phantom{ \overline{Q}^{A B} }
=
\widetilde{ {\mathbf{u}} } \; \widetilde{  {{D_{(1)}}} }  \sigma  \; \overline{\tau}^A
{{D_{(2)}}} \, {{D_{(2)}}}^{-1}
\,{\tau}^B \,  {{D_{(1)}}}
 {\psi_{(1)}}
$

$
\phantom{ \overline{Q}^{A B} }
=
\widetilde{ {\mathbf{u}} } \; \sigma  \;
\left(
 {{D_{(1)}}}^{-1} \; \overline{\tau}^A \,
{{D_{(2)}}}
\right)
\left({{D_{(2)}}}^{-1}\,{\tau}^B \,  {{D_{(1)}}}\right)
  {\psi_{(1)}}
$

$
\phantom{ \overline{Q}^{A B} }
=
L^{A}_{\phantom{A} C}\;
L^{B}_{\phantom{A} D}\;
\widetilde{ {\mathbf{u}} } \; \sigma  \;
\overline{\tau}^C
\,{\tau}^D \;
 {\psi_{(1)}}
=
L^{A}_{\phantom{A} C}\;
L^{B}_{\phantom{A} D}\;
Q^{C D}
$,\\
which follows from  Eq.[\ref{Ltaubar}] and Eq.[\ref{Ltau}].
In summary, under the action of $\overline{S0(4,4; \mathbb{R})}$,
\begin{eqnarray}
\mathbf{u} \mapsto \overline{\mathbf{u}} &=& {D_{(1)}} \;  \mathbf{u}
\nonumber \\
\psi_{(1)} \mapsto \overline{\psi}_1 &=& {D_{(1)}} \;  \psi_{(1)}
\nonumber \\
\psi_{(2)} \mapsto \overline{\psi}_2 &=& {D_{(2)}} \;  \psi_{(2)}.
\nonumber \\
{Q}^{A} \mapsto \overline{Q}^{A} &=& L^{A}_{\phantom{A} B}\;{Q}^{B}
\nonumber \\
{Q}^{A B} \mapsto \overline{Q}^{A B} &=& L^{A}_{\phantom{A} C}\;L^{B}_{\phantom{A} D}\;Q^{C D}
=
\{ L Q \widetilde{L}\}^{A B}
\label{transf}
\end{eqnarray}


\section{Appendix 3: Irreducible Representation of the $\tau$ Matrices}
\label{app:tau}

%
We adopt a real irreducible $8 \times 8$ matrix representation of the tau
matrices (see the Appendix) in which
$\overline{\tau}^{8} =  \mathbb{I}_{8 \times 8}  = \tau^{8}$.
Then by Eq.[\ref{tau-tau}]
$\overline{\tau}^{A} = - \tau^{A}$ for A = 1, ... ,7. Hence, again
by  Eq.[\ref{tau-tau}],
$(\tau^{A})^2$ is equal to $- \mathbb{I}_{8 \times 8} $  for A = 1,2,3 and is equal to  $+ \mathbb{I}_{8 \times 8} $ for A = 4,5,6, 7, 8.

A particular irreducible representation of the tau matrices  is

\begin{eqnarray}
\label{eq:tauMatrices}
&&
\begin{array}{cc}
{\tau^1=}\left(
 \begin{array}{cccccccc}
0 & 0 & 0 & 0 & 0 & 0 & 0 & -1  \\
 0 & 0 & 0 & 0 & 0 & 0 & -1 & 0  \\
 0 & 0 & 0 & 0 & 0 & 1 & 0 & 0  \\
 0 & 0 & 0 & 0 & 1 & 0 & 0 & 0  \\
 0 & 0 & 0 & -1 & 0 & 0 & 0 & 0  \\
 0 & 0 & -1 & 0 & 0 & 0 & 0 & 0  \\
 0 & 1 & 0 & 0 & 0 & 0 & 0 & 0  \\
 1 & 0 & 0 & 0 & 0 & 0 & 0 & 0
 \end{array}
 \right)
 &
{\tau^2=}\left(
 \begin{array}{cccccccc}
0 & 0 & 0 & 0 & 0 & 0 & 1 & 0  \\
 0 & 0 & 0 & 0 & 0 & 0 & 0 & -1  \\
 0 & 0 & 0 & 0 & -1 & 0 & 0 & 0  \\
 0 & 0 & 0 & 0 & 0 & 1 & 0 & 0  \\
 0 & 0 & 1 & 0 & 0 & 0 & 0 & 0  \\
 0 & 0 & 0 & -1 & 0 & 0 & 0 & 0  \\
 -1 & 0 & 0 & 0 & 0 & 0 & 0 & 0  \\
 0 & 1 & 0 & 0 & 0 & 0 & 0 & 0
 \end{array}
 \right)
\\
{\tau^3=}\left(
 \begin{array}{cccccccc}
0 & 0 & 0 & 0 & 0 & -1 & 0 & 0  \\
 0 & 0 & 0 & 0 & 1 & 0 & 0 & 0  \\
 0 & 0 & 0 & 0 & 0 & 0 & 0 & -1  \\
 0 & 0 & 0 & 0 & 0 & 0 & 1 & 0  \\
 0 & -1 & 0 & 0 & 0 & 0 & 0 & 0  \\
 1 & 0 & 0 & 0 & 0 & 0 & 0 & 0  \\
 0 & 0 & 0 & -1 & 0 & 0 & 0 & 0  \\
 0 & 0 & 1 & 0 & 0 & 0 & 0 & 0
 \end{array}
 \right)
 &
{\tau^4=}\left(
 \begin{array}{cccccccc}
0 & 0 & 0 & 0 & 0 & -1 & 0 & 0  \\
 0 & 0 & 0 & 0 & 1 & 0 & 0 & 0  \\
 0 & 0 & 0 & 0 & 0 & 0 & 0 & 1  \\
 0 & 0 & 0 & 0 & 0 & 0 & -1 & 0  \\
 0 & 1 & 0 & 0 & 0 & 0 & 0 & 0  \\
 -1 & 0 & 0 & 0 & 0 & 0 & 0 & 0  \\
 0 & 0 & 0 & -1 & 0 & 0 & 0 & 0  \\
 0 & 0 & 1 & 0 & 0 & 0 & 0 & 0
 \end{array}
 \right)
\\
{\tau^5=}\left(
 \begin{array}{cccccccc}
0 & 0 & 0 & 0 & 0 & 0 & 1 & 0  \\
 0 & 0 & 0 & 0 & 0 & 0 & 0 & 1  \\
 0 & 0 & 0 & 0 & -1 & 0 & 0 & 0  \\
 0 & 0 & 0 & 0 & 0 & -1 & 0 & 0  \\
 0 & 0 & -1 & 0 & 0 & 0 & 0 & 0  \\
 0 & 0 & 0 & -1 & 0 & 0 & 0 & 0  \\
 1 & 0 & 0 & 0 & 0 & 0 & 0 & 0  \\
 0 & 1 & 0 & 0 & 0 & 0 & 0 & 0
 \end{array}
 \right)
 &
{\tau^6=}\left(
 \begin{array}{cccccccc}
0 & 0 & 0 & 0 & 0 & 0 & 0 & 1  \\
 0 & 0 & 0 & 0 & 0 & 0 & -1 & 0  \\
 0 & 0 & 0 & 0 & 0 & 1 & 0 & 0  \\
 0 & 0 & 0 & 0 & -1 & 0 & 0 & 0  \\
 0 & 0 & 0 & -1 & 0 & 0 & 0 & 0  \\
 0 & 0 & 1 & 0 & 0 & 0 & 0 & 0  \\
 0 & -1 & 0 & 0 & 0 & 0 & 0 & 0  \\
 1 & 0 & 0 & 0 & 0 & 0 & 0 & 0
 \end{array}
 \right)
\\
{\tau^7=}\left(
 \begin{array}{cccccccc}
-1 & 0 & 0 & 0 & 0 & 0 & 0 & 0  \\
 0 & -1 & 0 & 0 & 0 & 0 & 0 & 0  \\
 0 & 0 & -1 & 0 & 0 & 0 & 0 & 0  \\
 0 & 0 & 0 & -1 & 0 & 0 & 0 & 0  \\
 0 & 0 & 0 & 0 & 1 & 0 & 0 & 0  \\
 0 & 0 & 0 & 0 & 0 & 1 & 0 & 0  \\
 0 & 0 & 0 & 0 & 0 & 0 & 1 & 0  \\
 0 & 0 & 0 & 0 & 0 & 0 & 0 & 1
 \end{array}
 \right)
 &
{\tau^8=}\left(
 \begin{array}{cccccccc}
1 & 0 & 0 & 0 & 0 & 0 & 0 & 0  \\
 0 & 1 & 0 & 0 & 0 & 0 & 0 & 0  \\
 0 & 0 & 1 & 0 & 0 & 0 & 0 & 0  \\
 0 & 0 & 0 & 1 & 0 & 0 & 0 & 0  \\
 0 & 0 & 0 & 0 & 1 & 0 & 0 & 0  \\
 0 & 0 & 0 & 0 & 0 & 1 & 0 & 0  \\
 0 & 0 & 0 & 0 & 0 & 0 & 1 & 0  \\
 0 & 0 & 0 & 0 & 0 & 0 & 0 & 1
\end{array}
\right)
\end{array}
\nonumber\\
\end{eqnarray}

\bibliographystyle{plain}
\bibliography{NASH-ID_IOParxiv}


\end{document}